# The dynamic magnetic behaviors of the Blume-Capel Ising bilayer system

**Mehmet Ertaş**[1]

*Department of Physics, Erciyes University, 38039 Kayseri, Turkey*

## Abstract

The dynamic magnetic behaviors of the spin-1 Blume-Capel Ising bilayer system (BCIS) are studied in an oscillating external magnetic field on a two-layer square lattice by utilizing the mean field theory based on Glauber-type stochastic dynamics (DMFT). The dynamic equations describing the time-dependencies of the average magnetizations are obtained with the Master equation. The dynamic phases in this system are found by solving these dynamic equations. The temperature dependence of the dynamic order parameters is examined to characterize the nature (continuous or discontinuous) of the phase transitions and to obtain the dynamic phase transition points (DPT). The dynamic phase diagrams are shown for ferromagnetic / ferromagnetic, antiferromagnetic / antiferromagnetic, antiferromagnetic / antiferromagnetic interactions in the plane of the reduced temperature versus magnetic field amplitude and they display dynamic tricritical and reentrant behavior as well as the dynamic triple point.

## 1. Introduction

Research on magnetic bilayers has received much interest both from the experimental and theoretical point of view in recent years [1-10]. Experimentally, several works have been published on the properties of the magnetic bilayer such as FePt/Fe [11], FM/TbFe (FM=Fe, Py-permalloy, FeCo) [12], SiGe, SiC and GeC [13], Ni/Au [14], Fe/Ni and Fe/Co [15]. On the other hand, theoretically, the equilibrium behaviors of bilayer Ising systems consisting of single ions, namely spin-1/2, spin-1 etc., have been widely studied by utilizing the well-known theoretical methods used in equilibrium statistical physics, such as the mean field theory (MFT) [16, 17], the effective field theory (EFT) [18, 19], the exact recursion equation on the Bethe lattice [6, 7], the renormalization-group (RG) method [20, 21], and Monte Carlo (MC) simulations [22, 23].

      Despite these equilibrium studies, the nonequilibrium aspects of bilayer Ising systems consisting of single ions have not been as thoroughly explored. There have only been few investigations, to our knowledge, about the dynamical aspects of the bilayer Ising system. These works are related to the dynamic behavior of the spin-1/2 bilayer system and can be summarized as follows. By using dynamic MC simulations, the short-time process and scaling behavior of a two-layer Ising model was studied by Wu et al. [24]. Jang and co-workers [25, 26] investigated the dynamic phase behaviors of magnetic thin ferromagnetic Heisenberg films using a dynamic MC simulation. The dynamic behavior of the spin-1/2 Ising model on a two-layer square lattices under the presence of a time-dependent oscillating external magnetic

---

[1]Corresponding author
*E-mail address:* mehmetertas@erciyes.edu.tr

field was investigated using Glauber-type stochastic dynamics based on the mean field theory (DMFT) [27] and the effective field theory (DEFT) [28]. In these studies, the dynamic phase transition (DPT) temperatures and the dynamic phase diagrams of the spin-1/2 bilayer Ising system are calculated in detail. We should also mention that very recently, the dynamic magnetic behaviors of the mixed Ising bilayer system consisting of different spins, namely mixed spins (1, 3/2), (3/2, 2), (2, 5/2) [29-32], by using DMFT for FM/FM, AFM/FM, and AFM/AFM interactions were investigated on the bilayer square lattice.

Unfortunately, the dynamic magnetic behaviors of the spin-1 Blume-Capel Ising bilayer (BCIB) system have not been studied so far. Therefore, the purpose of the present paper is to study the dynamic behaviors of the spin-1 BCIB system in an oscillating external magnetic field within DMFT. The DMFT equations for average magnetizations are obtained with the Glauber-type stochastic dynamics. We investigate the time variation of average magnetizations to find the phases and the thermal behavior of dynamic magnetizations to obtain the dynamic phase transition points, as well as to characterize the nature of the dynamic phase transitions. Finally, the dynamic phase diagrams are presented for FM/FM, AFM/FM, and AFM/AFM interactions in the plane of the reduced temperature versus magnetic field amplitude.

It can be noted that the physics of equilibrium phase transitions is well understood, and it is well established that structures arising from different dynamics that obey detailed balance and respect the same conservation laws exhibit universal asymptotic large-scale features. However, the mechanisms behind nonequilibrium phase transitions are not very well known, and the dependence on the specific dynamics is still an open question [33]. Therefore, further efforts on these challenging time-dependent problems, especially calculating the DPT points and constructing the phase diagram, should promise to be rewarding in the future. Hence, the DPT in an oscillating external magnetic field has attracted considerable attention in recent literature, theoretically (see [34-39] and references therein) and analytically [40, 41]. Moreover, experimental evidence for the DPT has been reported in polyethylene naphthalate (PEN) nanocomposites, in amorphous ultrathin Co films, cuprite superconductors, [Co/Pt]$_3$ magnetic multilayers, etc. [42-45]. Moreover, O'Malley et al. [46] studied the DPT in ecological systems in a temporally varying environment (seasonal variations).

The layout of this work is as follows. In Section 2, the model and the formulation are given briefly. In Section 3, the detailed numerical results and discussion are presented. A summary and conclusion are given in Section 4.

## 2. Model and derivation of mean field dynamic equations

The two-layer square lattice is an extension of its one-layer version. Two identical layers, namely the upper layer ($G_1$) and lower layer ($G_2$) of the square lattices, are placed parallel to each other, forming the two-layer square lattice. The schematic representation of the system is illustrated in Fig. 1. Each layer has N sites and interacts with its nearest-neighbor (NN) and the corresponding adjacent spins in the other layer whose sites are labeled by i, i′, j and j′, as seen in Fig. 1. Such a system may be described by the following Hamiltonian:

$$\mathcal{H} = -J_1 \sum_{<i'j'>} \sigma_{i'}\sigma_{j'} - J_2 \sum_{<ij>} S_i S_j - J_3 \left( \sum_{<ii'>} \sigma_{i'} S_i + \sum_{<jj'>} \sigma_{j'} S_j \right) - D \left( \sum_{<i'>} \sigma_{i'}^2 + \sum_{<j'>} \sigma_{j'}^2 + \sum_{<i>} S_i^2 + \sum_{<j>} S_j^2 \right)$$

$$- H \left( \sum_{<i'>} \sigma_{i'} + \sum_{<j'>} \sigma_{j'} + \sum_{<i>} S_i + \sum_{<j>} S_j \right),$$

(1)

where <ij> and <i'j'> indicate a summation over all the pairs of nearest-neighboring sites of each layer. $J_1$ and $J_2$ denote exchange constants for the first and second layer and are also called intralayer coupling constants. $J_3$ is the interlayer coupling constant over all the adjacent neighboring sites of layers, as seen in Fig. 1. D is the single ion anisotropy constant and H is a time-dependent oscillating external magnetic field: $H(t)=H_0 \cos(wt)$, where $H_0$ and $w = 2\pi\nu$ are the amplitude and the angular frequency of the oscillating field, respectively. The system is in contact with an isothermal heat bath at absolute temperature $T_A$.

On the other hand, for the formulation of the system, one needs to introduce the order parameters, namely magnetizations. From Fig. 1, one can see that each layer of the system is also a two-lattice system with spin variables $\sigma_i = \pm 1, 0$ and $\sigma_{i'} = \pm 1, 0$ on the sites of the sublattices A and B, respectively, and $S_j = \pm 1, 0$ and $S_{j'} = \pm 1, 0$ on the sites of the sublattices A and B, respectively. Therefore, the system can be described with four sublattice magnetizations or four simple magnetizations. These magnetizations are introduced as follows: $m_1^A \equiv \langle \sigma_i \rangle$, $m_1^B \equiv \langle \sigma_j \rangle$, $m_2^A \equiv \langle S_{i'} \rangle$, $m_2^B \equiv \langle S_{j'} \rangle$, where $\langle \ \rangle$ is the thermal expectation value.

Now, by utilizing Glauber-type stochastic dynamics, the dynamic mean field equations can be obtained. In particular, we employ Glauber transition rates to obtain the set of dynamic mean field equations. The system evolves according to a Glauber-type stochastic process at a rate of $1/\tau$ transitions per unit time; hence the frequency of spin flipping, $f$, is $1/\tau$. $P(\sigma_1, \sigma_2, ..., \sigma_N; t)$ and $P(S_1, S_2, ..., S_N; t)$ are the probability functions when the system owns $\sigma_1, \sigma_2, ..., \sigma_N$ and $S_1, S_2, ..., S_N$ spin configurations at time t. If we let $W_i(\sigma_i^A \to \sigma_i^{A'})$ be the probability per unit time that the *i*th spin changes from the value $\sigma_i^A$ to $\sigma_i^{A'}$, while the S spins on the sublattice B are momentarily fixed, then we may write the time derivate of the $P_1^A(\sigma_1^A, \sigma_2^A, ..., \sigma_N^A; t)$ as

$$\frac{d}{dt} P_1^A(\sigma_1^A, \sigma_2^A, ..., \sigma_N^A; t) = -\sum_i \left( \sum_{\sigma_i^A \neq \sigma_i^{A'}} W_i^A(\sigma_i^A \to \sigma_i^{A'}) \right) P_1^A(\sigma_1^A, \sigma_2^A, ..., \sigma_i^A, ..., \sigma_N^A; t)$$
$$+ \sum_i \left( \sum_{\sigma_i^A \neq \sigma_i^{A'}} W_i^A(\sigma_i^{A'} \to \sigma_i^A) P_1^A(\sigma_1^A, \sigma_2^A, ..., \sigma_i^{A'}, ..., \sigma_N^A; t) \right). \tag{2}$$

This equation is called the Master equation. $W_i^A(\sigma_i^A \to \sigma_i^{A'})$ is the probability per unit time that the *i*th spin changes from the value $\sigma_i^A$ to $\sigma_i^{A'}$. In this sense the Glauber model is stochastic. Since the system is in contact with a heat bath at absolute temperature $T_A$, each spin can change from the value $\sigma_i^A$ to $\sigma_i^{A'}$ with the probability per unit time

$$W_i^A(\sigma_i^A \to \sigma_i^{A'}) = \frac{1}{\tau} \frac{\exp(-\beta \Delta E(\sigma_i^A \to \sigma_i^{A'}))}{\sum_{\sigma_i^{A'}} \exp(-\beta \Delta E(\sigma_i^A \to \sigma_i^{A'}))}, \tag{3}$$

where $\sum_{\sigma_i^{A'}}$ is the sum over the two possible values of $\sigma_i^{A'} = \pm 1, 0$, and

$$\Delta E_i^A (\sigma_i^A \to \sigma_i^{A'}) = -(\sigma_i^{A'} - \sigma_i^A)(J_1 \sum_j \sigma_j^B + J_3 \sum_{i'} S_{i'}^A + H), \tag{4}$$

gives the change in the energy of the system when the $\sigma_i$-spin changes. The probabilities satisfy the detailed balance condition

$$\frac{W_i^A(\sigma_i^A \to \sigma_i^{A'})}{W_i^A(\sigma_i^{A'} \to \sigma_i^A)} = \frac{P(\sigma_1^A, \sigma_2^A, ..., \sigma_i^{A'}, ..., \sigma_N^A)}{P(\sigma_1^A, \sigma_2^A, ..., \sigma_i^A, ..., \sigma_N^A)}, \tag{5}$$

and by substituting the possible values of $\sigma_i$, we get

$$W_i^A(1 \to 0) = W_i^A(-1 \to 0) = \frac{1}{\tau} \frac{\exp(-\beta D)}{2\cosh(\beta x) + \exp(-\beta D)}, \tag{6a}$$

$$W_i^A(1 \to -1) = W_i^A(0 \to -1) = \frac{1}{\tau} \frac{\exp(-\beta x)}{2\cosh(\beta x) + \exp(-\beta D)}, \tag{6b}$$

$$W_i^A(0 \to 1) = W_i^A(-1 \to 1) = \frac{1}{\tau} \frac{\exp(\beta x)}{2\cosh(\beta x) + \exp(-\beta D)}, \tag{6c}$$

where $x = J_1 \sum_{\langle j' \rangle} \sigma_{j'}^B + J_3 \sum_i S_i^A + H$. From the Master equation associated with the stochastic process, it follows that the average $\langle \sigma_k^A \rangle$ satisfies the following equation:

$$\tau \frac{d}{dt} \langle \sigma_k^A \rangle = -\langle \sigma_k^A \rangle + \left\langle \frac{2\sinh\left[\beta\left(J_1 \sum_{j'} \sigma_{j'}^B + J_3 \sum_i S_i^A + H\right)\right] + 2\exp(4\beta D)\sinh\left[\beta\left(J_1 \sum_{j'} \sigma_{j'}^B + J_3 \sum_i S_i^A + H\right)\right]}{2\cosh\left[\beta\left(J_1 \sum_{j'} \sigma_{j'}^B + J_3 \sum_i S_i^A + H\right)\right] + \exp(-\beta D)} \right\rangle.$$

(7)

In terms of the mean field approach, this dynamic equation can be written as follows:

$$\Omega \frac{d}{d\xi} m_1^A = -m_1^A + \frac{2\sinh\left[\frac{2}{T}\left(zm_1^B + \frac{J_3}{J_1} m_2^A + h\cos\xi\right)\right] + \exp\left(\frac{d}{T}\right)\sinh\left[\frac{1}{T}\left(zm_1^B + \frac{J_3}{J_1} m_2^A + h\cos\xi\right)\right]}{2\cosh\left[\frac{2}{T}\left(zm_1^B + \frac{J_3}{J_1} m_2^A + h\cos\xi\right)\right] + \exp\left(-\frac{d}{T}\right)},$$

(8)

where $m_1^A \equiv \langle \sigma_i^A \rangle$, $m_1^B \equiv \langle \sigma_{j'}^B \rangle$, $m_2^A \equiv \langle S_i^A \rangle$, $m_2^B \equiv \langle S_j^B \rangle$, $\xi = wt$, $T = (\beta J_1)^{-1}$, $h = H_0/J_1$, $d = D/J_1$ and $\Omega = \tau w$. We fixed $z = 4$ and $w = 2\pi\nu$.

The other dynamic equations concerning the $G_1$ and $G_2$ layers can be similarly obtained as follows:

$$\Omega \frac{d}{d\xi} m_1^B = -m_1^B + \frac{2\sinh\left[\frac{2}{T}\left(zm_1^A + \frac{J_3}{J_1}m_2^B + h\cos\xi\right)\right] + \exp\left(\frac{d}{T}\right)\sinh\left[\frac{1}{T}\left(zm_1^A + \frac{J_3}{J_1}m_2^B + h\cos\xi\right)\right]}{2\cosh\left[\frac{2}{T}\left(zm_1^A + \frac{J_3}{J_1}m_2^B + h\cos\xi\right)\right] + \exp\left(-\frac{d}{T}\right)}, \quad (9)$$

$$\Omega \frac{d}{d\xi} m_2^A = -m_2^A + \frac{2\sinh\left[\frac{2}{T}\left(\frac{J_2}{J_1}zm_2^B + \frac{J_3}{J_1}m_1^A + h\cos\xi\right)\right] + \exp\left(\frac{d}{T}\right)\sinh\left[\frac{1}{T}\left(\frac{J_2}{J_1}zm_2^B + \frac{J_3}{J_1}m_1^A + h\cos\xi\right)\right]}{2\cosh\left[\frac{2}{T}\left(\frac{J_2}{J_1}zm_2^B + \frac{J_3}{J_1}m_1^A + h\cos\xi\right)\right] + \exp\left(-\frac{d}{T}\right)},$$

(10)

$$\Omega \frac{d}{d\xi} m_2^B = -m_2^B + \frac{2\sinh\left[\frac{2}{T}\left(\frac{J_2}{J_1}zm_2^A + \frac{J_3}{J_1}m_1^B + h\cos\xi\right)\right] + \exp\left(\frac{d}{T}\right)\sinh\left[\frac{1}{T}\left(\frac{J_2}{J_1}zm_2^A + \frac{J_3}{J_1}m_1^B + h\cos\xi\right)\right]}{2\cosh\left[\frac{2}{T}\left(\frac{J_2}{J_1}zm_2^A + \frac{J_3}{J_1}m_1^B + h\cos\xi\right)\right] + \exp\left(-\frac{d}{T}\right)}.$$

(11)

Thus a set of mean-field dynamical equations are obtained.

## 3. Numerical results and discussions

### 3.1. Phases in the system

We solved Eqs. (8)-(11) look for steady states then classified their behavior to find the phases as paramagnetic (p), ferromagnetic (f), antiferromagnetic (af), surface (sf), compensated (c), mixed (m) and non-magnetic (nm). The stationary solutions of Eqs. (8) and (11) will be a periodic function of $\xi$ with period $2\pi$; that is $m^{A, B}(\xi+2\pi) = m^{A, B}(\xi)$. Moreover, they can be one of three types according to whether they have or do not have the property

$$m_1^A(\xi+\pi) = -m_1^A(\xi) \text{ and } m_1^B(\xi+\pi) = -m_1^B(\xi), \quad (12a)$$

and

$$m_2^A(\xi+\pi) = -m_2^A(\xi) \text{ and } m_2^B(\xi+\pi) = -m_2^B(\xi). \quad (12b)$$

A solution satisfying Eqs. (12a) and (12b) is called a symmetric solution which corresponds to a paramagnetic (p) solution and it exists at high values of T and h. In this solution, the average magnetizations are equal to each other. They oscillate around the zero value and are delayed with respect to the external magnetic field. An example is shown in Fig. 2(a). The second type of solution, which does not satisfy Eqs. (12a) and (12b), is called a nonsymmetric solution and magnetizations do not follow the external magnetic field anymore; instead of oscillating around the zero value, they oscillate around the nonzero value. These fundamental phases are defined as follows. (i) The ferromagnetic (f) phase: $m_1^A = m_1^B \neq 0$ with positive spin values and $m_2^A = m_2^B \neq 0$ and with positive spin values. (ii) The antiferromagnetic (af) phase: $m_1^A = -m_1^B$, $-m_2^A = m_2^B$. (iii) The surface ferromagnetic (sf) phase: $m_1^A = m_1^B \neq 0$, $-m_2^A = m_2^B \neq 0$. (iv) The compensated (c) phase: $m_1^A = m_1^B \neq 0$ with positive spin values and $m_2^A = m_2^B \neq 0$ with negative spin values, or $m_1^A = m_1^B \neq 0$ with negative spin values and $m_2^A = m_2^B \neq 0$ with positive spin values. (v) The mixed (m) phase: $m_1^A = -m_1^B$, $m_2^A = -m_2^B$. These phases are defined according to Refs. [27, 31, 32] and are illustrated in Fig. 2(b)-(f). These phases can be better understood by examination of Table 1. In addition to these seven fundamental phases, seven coexistence solutions or mixed phases are found, which are composed of binary combinations of the fundamental phases, namely the f + p, f + c, f + m, sf + nm, nm + p, m + p and af + p.

*3.2. Behaviors of dynamic magnetizations and dynamic phase transition points*

The dynamic magnetizations ($M_1^{A,B}$ and $M_2^{A,B}$), are defined as

$$M_1^A = \frac{1}{2\pi}\int_0^{2\pi} m_1^A(\xi)d\xi, \quad M_1^B = \frac{1}{2\pi}\int_0^{2\pi} m_1^B(\xi)d\xi, \qquad (13)$$

and

$$M_2^A = \frac{1}{2\pi}\int_0^{2\pi} m_2^A(\xi)d\xi, \quad M_2^B = \frac{1}{2\pi}\int_0^{2\pi} m_2^B(\xi)d\xi. \qquad (14)$$

In order to determine the dynamic phase transition (DPT) temperatures among the phases we will study the temperature dependences of the dynamic magnetizations, by solving Eqs. (13)-(14) numerically. This investigation also leads us to characterize the nature (first- or second-order) of the DPT temperatures. A few interesting results are plotted in Figs. 3(a)-(f). Respectively, $T_C$ and $T_t$ denotes second-and first-order phase transition temperature. Fig. 3(a) is calculated for $J_1=1.0$, $J_2/|J_1|=1.0$, $J_3/|J_1|=1.0$, d = 1.0, h = 0.5 and illustrates a second-order phase transition which is from the f phase to the p phase at $T_C = 3.63$. Because, at zero value of temperature, $M_1^{A,B} = M_2^{A,B} = 1.0$ and as the temperature increases, they reduce to zero value incessantly. Fig. 3(b) is obtained for $J_1=1.0$, $J_2/|J_1|=1.0$, $J_3/|J_1|=1.0$, d = 1.0, h = 3.77 and it shows first-order phase transition which is from the f phase to the p phase at $T_t = 0.75$. Since, at zero value of temperature, $M_1^{A,B} = M_2^{A,B} = 1.0$ and as the temperature increases, they decrease to zero value transient by the time $T_t$. Figs. 3(c) and 3(d) are plotted for $J_1=1.0$, $J_2/|J_1|=1.0$, $J_3/|J_1|=1.0$, d = -1.0, h = 0.1 and various different initial values. The behavior of Fig. 3(c), is similar to that of Fig. 3(a); hence the system undergoes a second-order phase transition from the f phase to the p phase at $T_C= 2.95$. In Fig. 3(d), the system undergoes two successive phase transitions. The first is a first-order transition, because discontinuity occurs at $T_t = 0.28$. The transition is from the p phase to the f phase. The second is a second-order transition from the f phase to the p phase at $T_C = 2.95$. This means that the coexistence region,

i.e the f + p mixed phase, exists in the system. Figs. 3(e) and 3(f) are obtained for $J_1 = -1.0$, $J_1 = -2.0$, $J_3 = -3.0$, d = -1.0, h = 0.1 and various different initial values. In Fig. 3(e), $M_1^A = M_2^B = 1$ and $M_1^B = M_2^A$ at the zero temperature and they decrease to zero continuously as the temperature increases; hence the system undergoes a second-order phase transition from the af phase to the p phase at $T_C$ = 4.30. In Fig. 3(f), $M_1^{A,B} = M_2^{A,B}$ is always equal to zero; hence the system does not undergo any phase transition. This figure corresponds to the p phase. From Figs. 3(e) and 3(f), one can see that the af + p mixed phase region exists until $T_C$. It is worthwhile mentioning that two successive transitions have also been experimentally observed in $DyVO_4$ [47].

### 3.3. Dynamic phase diagrams

Having obtained the dynamic phase transition (DPT) points, we are now ready to investigate the dynamic phase diagrams (DPD). Figs. 4-6 illustrate the FM / FM, AFM / FM, AFM / AFM interactions, respectively. In these figures solid and dashed lines represent the second-order ($T_C$-lines) and first-order ($T_t$-lines) phase transition lines and the filled circle (•) corresponds to the dynamic tricritical point.

#### 3.3.1. The case of the FM/FM interaction

The dynamic phase diagrams of the BCIB system with the case of FM/FM interactions for $J_1 = 1.0$, $J_2/|J_1| = 1.0$, $J_3/|J_1| = 1.0$ and various values of d are illustrated in Figs. 4 (a)-(e), namely d= 1.0, -1.0, -2.0, -2.5, -3.0, respectively. In Fig. 4, the following five interesting phenomena are observed. (i) The phase diagrams show one, two and three dynamic tricritical points that separate a second-order transition line from a first-order transition line, seen in Figs. 4(a)-(c), (e) and (d). (ii) The tp dynamic special point, at which three first-order transition lines meet, occurs in Figs. 4(a), (b), (d). (iii) Figs. 4(a), (b) and (d) display both a tricritical point and a triple point (TP), but Figs. 4(c) and (e) show only a tricritical point. (iv) The BCIB system always illustrates the ferromagnetic fundamental phase. (v) The mixed phases usually separate with first-order phase transition lines from the fundamental phases. On the other hand, the dynamic phase boundaries among the fundamental phases are usually second-order phase transition lines.

#### 3.3.2. The case of the AFM/FM interaction

The dynamic phase diagrams of the BCIB system with the case of AFM/FM interactions for $J_1 = -1.0$, $J_2/|J_1| = 1.0$, $J_3/|J_1| = 0.1$ and various values of d are illustrated in Figs. 5 (a)-(e), namely d= 1.0, -1.0, -2.0, -2.5, -3.0, respectively. These phase diagrams in Fig. 5 are similar to Fig. 4, and from these phase diagrams the following four interesting phenomena were observed. (i) Five different phase diagrams are obtained for AFM/FM interactions. (ii) The BCIB system includes the p and nm fundamental phases as well as the sf + nm, nm + p and nm + p mixed phases. (iii) While the mixed phases usually separate with first-order phase transition lines from the fundamental phases, the dynamic phase boundaries among the fundamental phases are usually second-order phase transition lines. (iv) The phase diagrams display one, and two dynamic tricritical points, as seen in Figs. 5(a), 5(b) and 5(c)-(e). (v) Only Fig. 5(c) illustrates the dynamic special point, namely tp.

#### 3.3.3. The case of the AFM/AFM interaction

The dynamic phase diagram of the BCIB system with the case AFM/AFM The phase diagrams in Fig. 6 are not similar to those in Figs. 4 and 5. From these phase diagrams, the following four interesting phenomena were observed. (i) Five different phase diagrams are obtained for AFM/AFM interactions. (ii) These dynamic phase diagrams, namely Fig. 6(a), (b) and (d), show a reentrant behavior. (iii) The BCIB system contains the p fundamental phase as well as the f+ nm, m + p and af + p mixed phases. (iv) One, two or three dynamic tricritical points are seen in these phase diagrams. (v) The dynamic phase diagrams do not contain a dynamic special point.

It has long been recognized that similar phase diagrams to those in Figs. 4 (a), 4(b), 5(a), 5(b), 6(d) and 6(e) have also been obtained in the kinetic mixed (1, 2) and mixed (2, 5/2) systems (see [31, 32, 48] and references therein). On the other hand, the dynamic phase diagrams in Figs. 4(c)–(e), 5(c)-(e), 6(a)-(c) are only observed in the BCIB system; hence they were not obtained by previous studies of the kinetic Ising (see [31, 32, 48] and references therein).

## 4. Summary and Conclusion

By utilizing the mean field theory based on Glauber-type stochastic dynamics (DMFT), the dynamic magnetic behaviors of the BCIS are investigated under the presence of a time-dependent oscillating external magnetic field. Glauber-type stochastic dynamics were used to describe the time evolution of the system and the time variations of the average order parameters were studied in order to find the phases in the systems. Then, as a function of the reduced temperature, the behavior of the average order parameters in a period or the dynamic order parameters was investigated; this study led us to characterize the (continuous or discontinuous) nature of the dynamic phase transitions and to obtain the dynamic phase transition (DPT) points. Finally, the dynamic phase diagrams were presented for ferromagnetic / ferromagnetic, antiferromagnetic / antiferromagnetic, antiferromagnetic / antiferromagnetic interactions in the plane of the reduced temperature versus magnetic field amplitude and they display dynamic tricritical and reentrant behavior as well as the dynamic triple point.

We hope that our detailed theoretical investigation may stimulate further works to study the DPT and the dynamic hysteresis in the mixed Ising model using more accurate techniques, such as kinetic Monte Carlo (MC) simulations or renormalization-group (RG) calculations, and that it may also be of assistance in further experimental research. Moreover, it may stimulate theoretical and experimental research to explore more complicated systems.

**List of figure captions**

**Fig. 1.** Two-layer square lattice, $G_1$ and $G_2$ refer to the upper and lower layers containing the spins labeled as $\sigma_{i'}$, $\sigma_{j'}$ and $S_i$, $S_j$.

**Fig. 2.** Time variations of the magnetizations ($m_1^A$, $m_1^B$, $m_2^A$ and $m_2^B$):

a) Exhibiting paramagnetic phase (p), $J_1 = 1.0$, $J_2/|J_1|=1.0$, $J_3/|J_1|=1.0$, $h = 5.0$, $d = 1.0$ and $T = 2.0$;
b) Exhibiting ferromagnetic phase (f), $J_1 = 1.0$, $J_2/|J_1|=1.0$, $J_3/|J_1|=1.0$, $h = 1.0$, $d = 1.0$ and $T = 2.0$;
c) Exhibiting antiferromagnetic phase (af), $J_1 = -1.0$, $J_2/|J_1|=-1.0$, $J_3/|J_1|=-3.0$, $h = 3.0$, $d = -1.0$ and $T = 1.52$;
d) Exhibiting surface phase (sf), $J_1 = -1.0$, $J_2/|J_1|=1.0$, $J_3/|J_1|=0.1$, $h = 1.5$, $d = 1.0$ and $T = 1.0$;
e) Exhibiting compensated phase (c), $J_1 = 1.0$, $J_2/|J_1|=1.0$, $J_3/|J_1|=1.0$, $h = 2.05$, $d = -2.0$ and $T = 0.05$;
f) Exhibiting mixed phase (m), $J_1 = -1.0$, $J_2/|J_1|=-1.0$, $J_3/|J_1|=5.0$, $h = 4.0$, $d = 0.1$ and $T = 3.0$;
g) Exhibiting nonmagnetic phase (nm), $J_1 = -1.0$, $J_2/|J_1|=0.5$, $J_3/|J_1|=0.1$, $h = 3$, $d = 0.1$ and $T = 2$.

**Fig. 3.** Reduced temperature dependences of dynamic magnetizations $M_1^A$, $M_1^B$ and $M_2^A$, $M_2^B$. $T_t$ and $T_C$ are the critical or the first-order phase transition and the second-order phase transition temperatures, respectively.

a) Exhibiting second-order phase transition from the f phase to the p phase for $J_1 = 1.0$, $J_2/|J_1|=1.0$, $J_3/|J_1|=1.0$, $d = 1.0$ and $h = 0.5$; $T_C$ is found at 3.63.

b) Exhibiting first-order phase transition from the f phase to the p phase for $J_1 = 1.0$, $J_2/|J_1|=1.0$, $J_3/|J_1|=1.0$, $d = 1.0$ and $h = 3.77$; $T_t$ is found at 0.75.

c) and d) Exhibiting two successive phase transitions, the first is a first-order phase transition from the f + p phase to the f phase and the second one is a second-order phase transition from the f to the p phase for $J_1 = 1.0$, $J_2/|J_1|=1.0$, $J_3/|J_1|=1.0$, $d = -1.0$ and $h = 0.1$; 0.28 and 2.95 are found for $T_t$ and $T_C$, respectively.

e) and f) Exhibiting second-order phase transitions from the af + p phase to the p phase for $J_1 = -1.0$, $J_1 = -2.0$, $J_3 = -3.0$, $d = -1.0$ and $h = 0.1$; 4.3 is found for $T_C$.

**Fig. 4.** Phase diagrams of spin-1 Blume-Capel Ising bilayer (BCIB) system on two-layer square lattice in (T, h) plane for FM/FM, i.e. $J_1 > 0$ and $J_2 > 0$. Dashed lines represent first-order phase transition, and tp represents the dynamic triple point. For $J_1 = 1.0$, $J_2/|J_1|=1.0$, $J_3/|J_1|=1.0$ and **a)** $d = 1.0$; **b)** $d = -1.0$; **c)** $d = -2.0$, **d)** $d = -2.5$ and **e)** $d = -3.0$.

**Fig. 5.** Fig. 5 same as Fig. 4, phase diagrams of BCIB system on two-layer square lattice in (T, h) plane for AFM/FM, i.e. $J_1 < 0$ and $J_2 > 0$. For $J_1 = -1.0$, $J_2/|J_1|=1.0$, $J_3/|J_1|=0.1$ and **a)** $d = 1.0$; **b)** $d = -1.0$; **c)** $d = -2.0$, **d)** $d = -2.5$ and **e)** $d = -3.0$.

**Fig. 6.** Fig. 6 same as Fig. 4, phase diagrams of BCIB system on two-layer square lattice in (T, h) plane for AFM/AFM, i.e. $J_1 < 0$ and $J_2 < 0$. **a)** $J_1 = -1.0$, $J_2/|J_1|=-1.0$, $J_3/|J_1|=5.0$, $d =$

0.1; **b)** $J_1 = -1.0$, $J_2/|J_1| = -1.0$, $J_3/|J_1| = 5.0$, d = 1.0; **c)** $J_1 = -1.0$, $J_2/|J_1| = -1.0$, $J_3/|J_1| = 5.0$, d = -1.0 **d)** $J_1 = -1.0$, $J_2/|J_1| = -1.0$, $J_3/|J_1| = 5.0$, d = -2.0 and **e)** $J_1 = -1.0$, $J_2/|J_1| = -1.0$, $J_3/|J_1| = -3.0$, d = -1.0

**Table caption**
**Table 1.** Characteristics of time variations of magnetizations ( $m_1^A(\xi)$, $m_1^B(\xi)$, $m_2^A(\xi)$, $m_2^B(\xi)$ ).

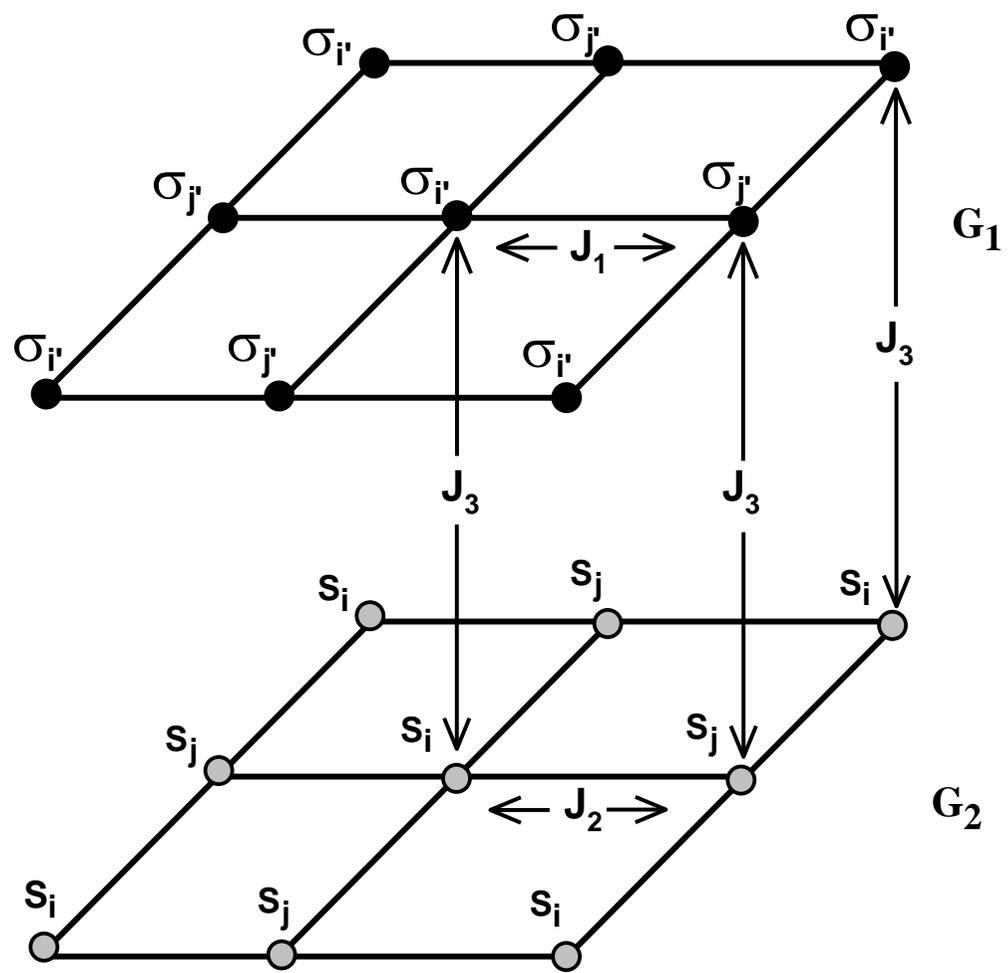

**Fig. 1**

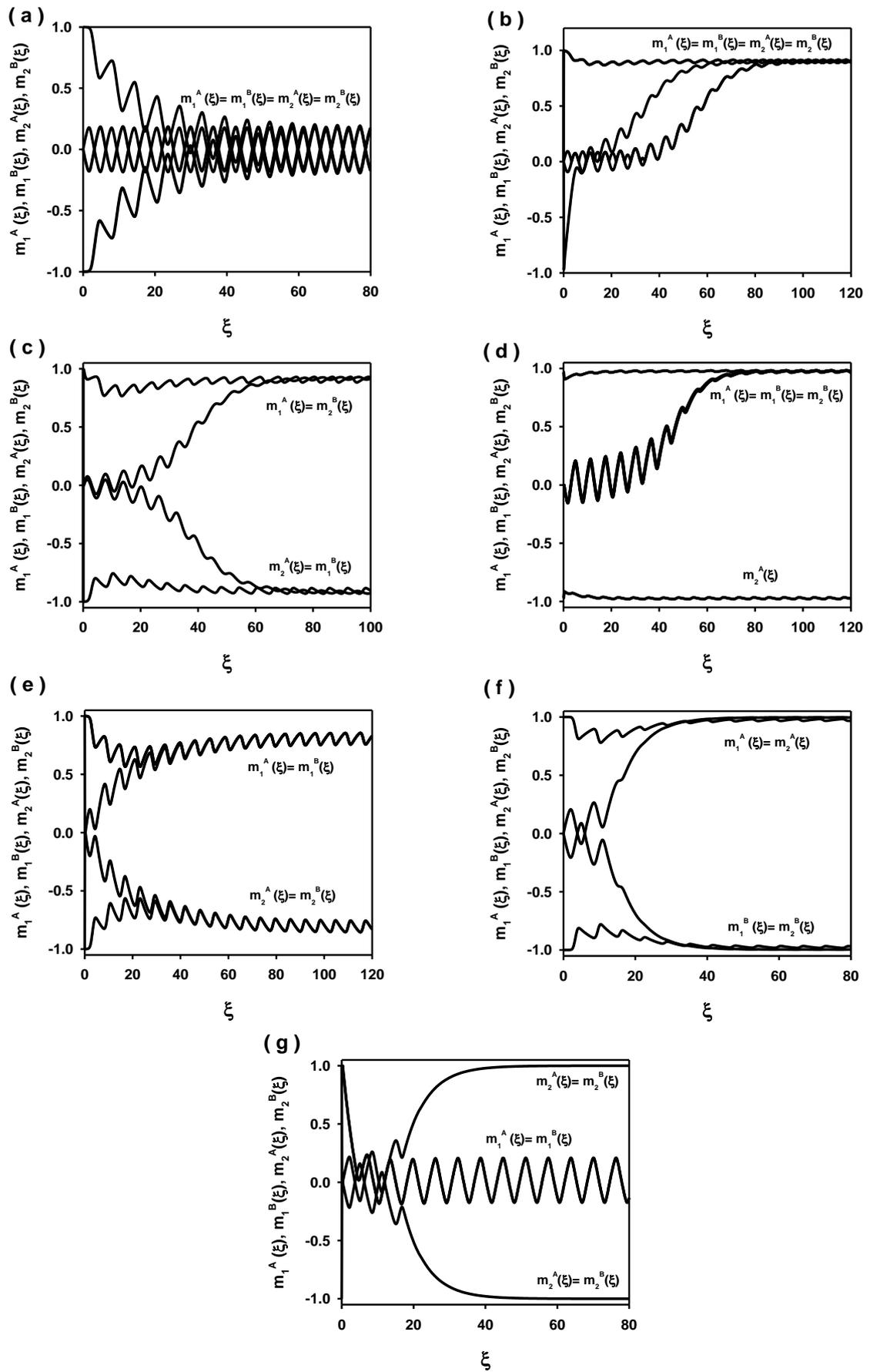

**Fig. 2**

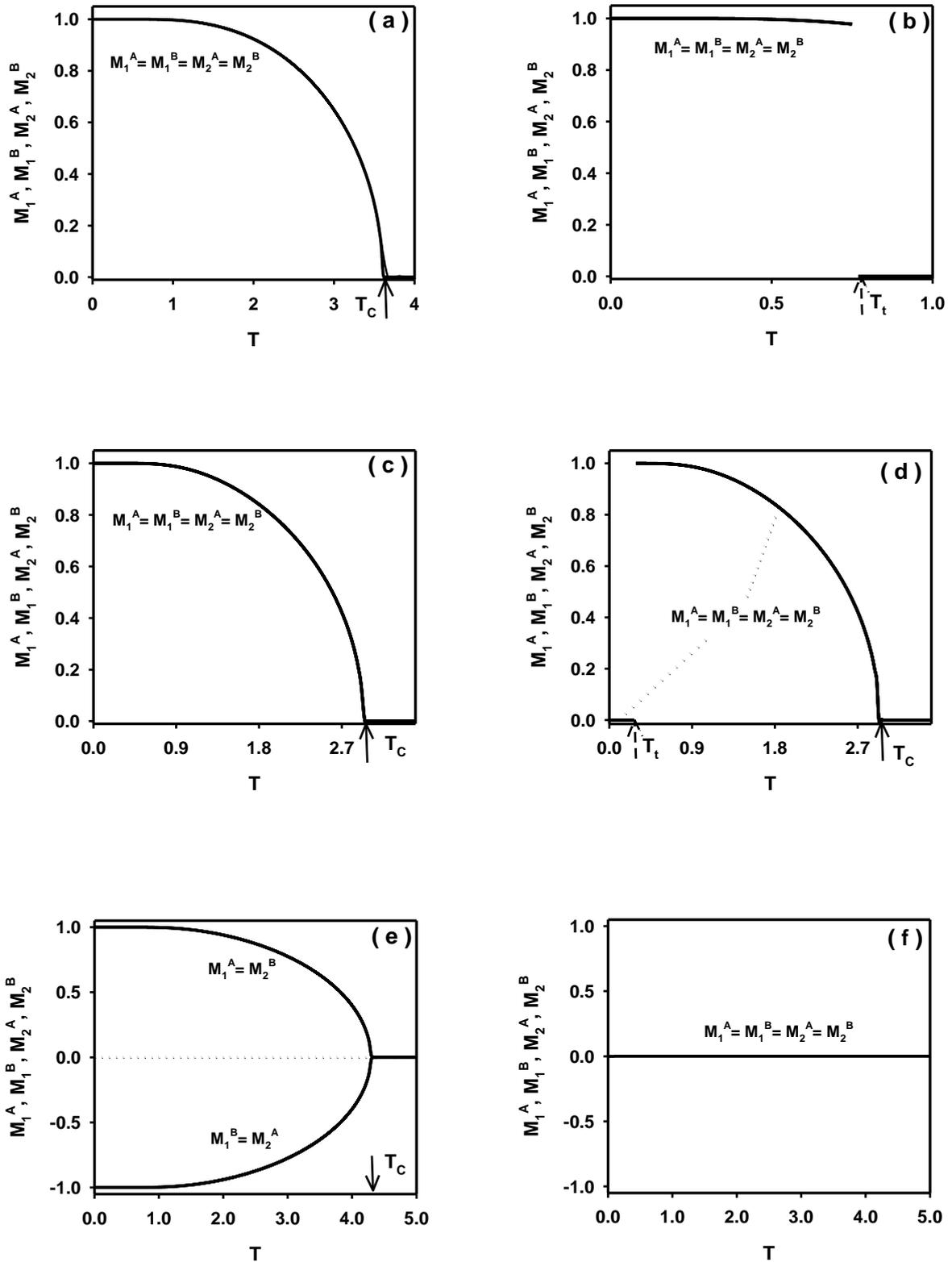

**Fig. 3**

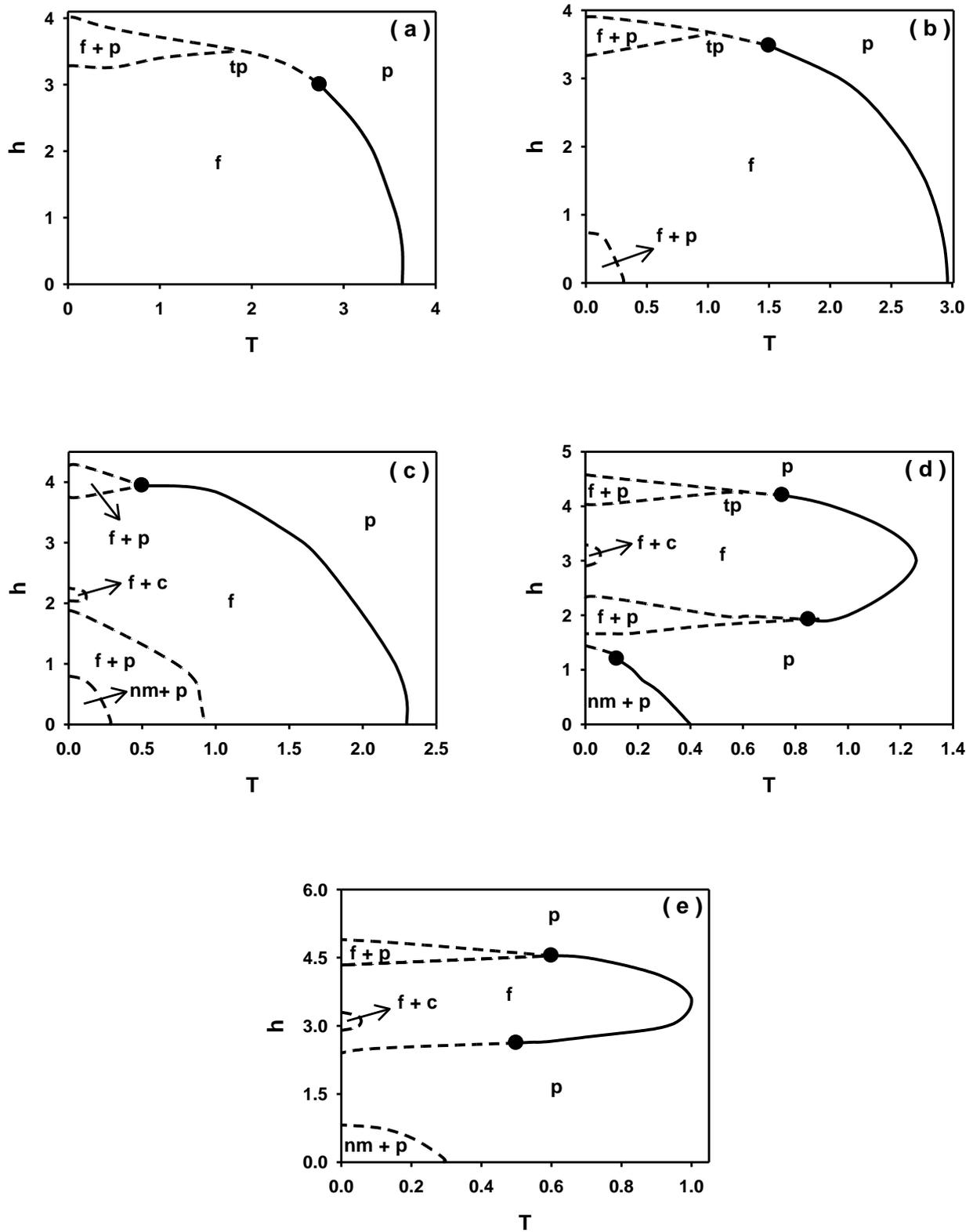

**Fig. 4**

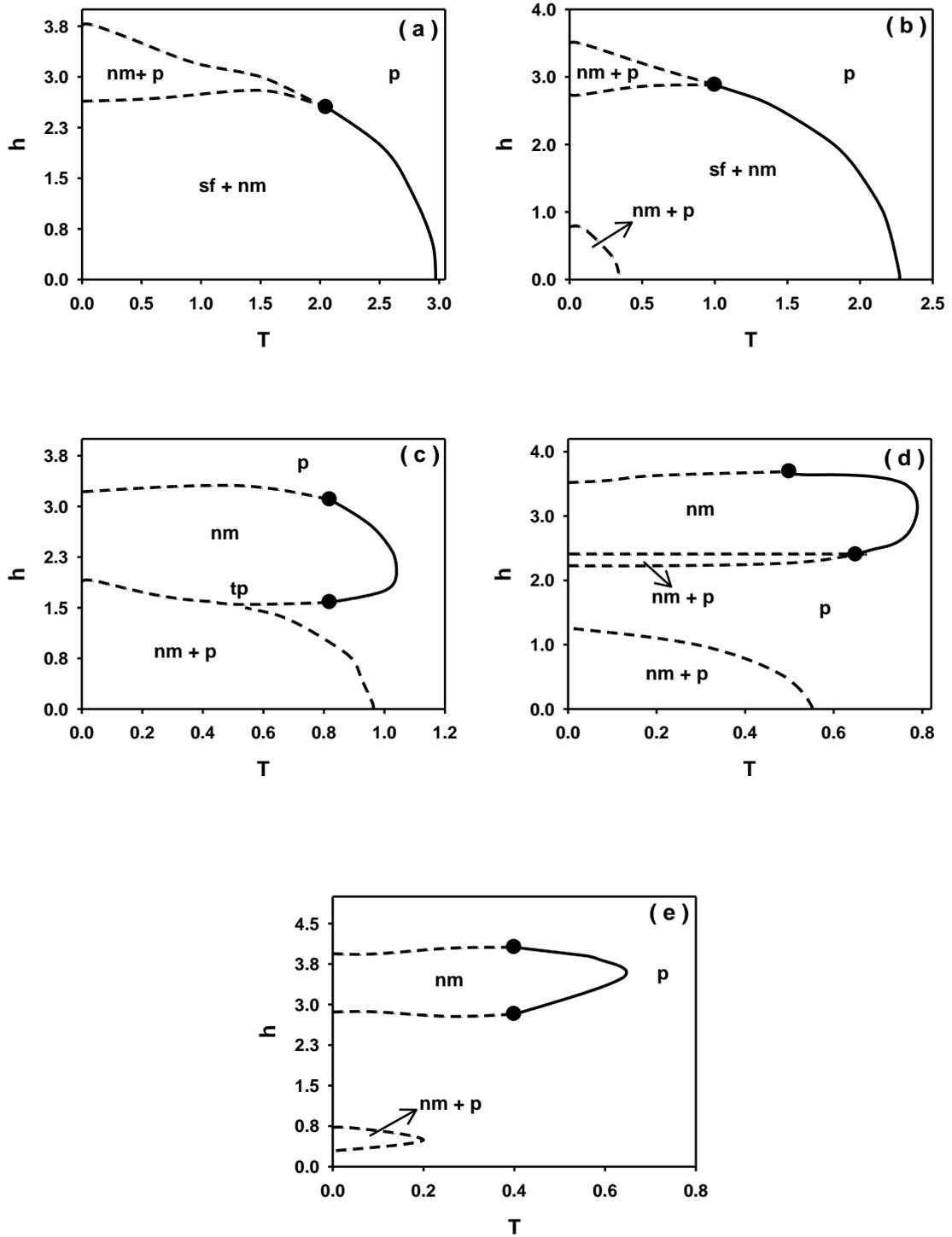

**Fig. 5**

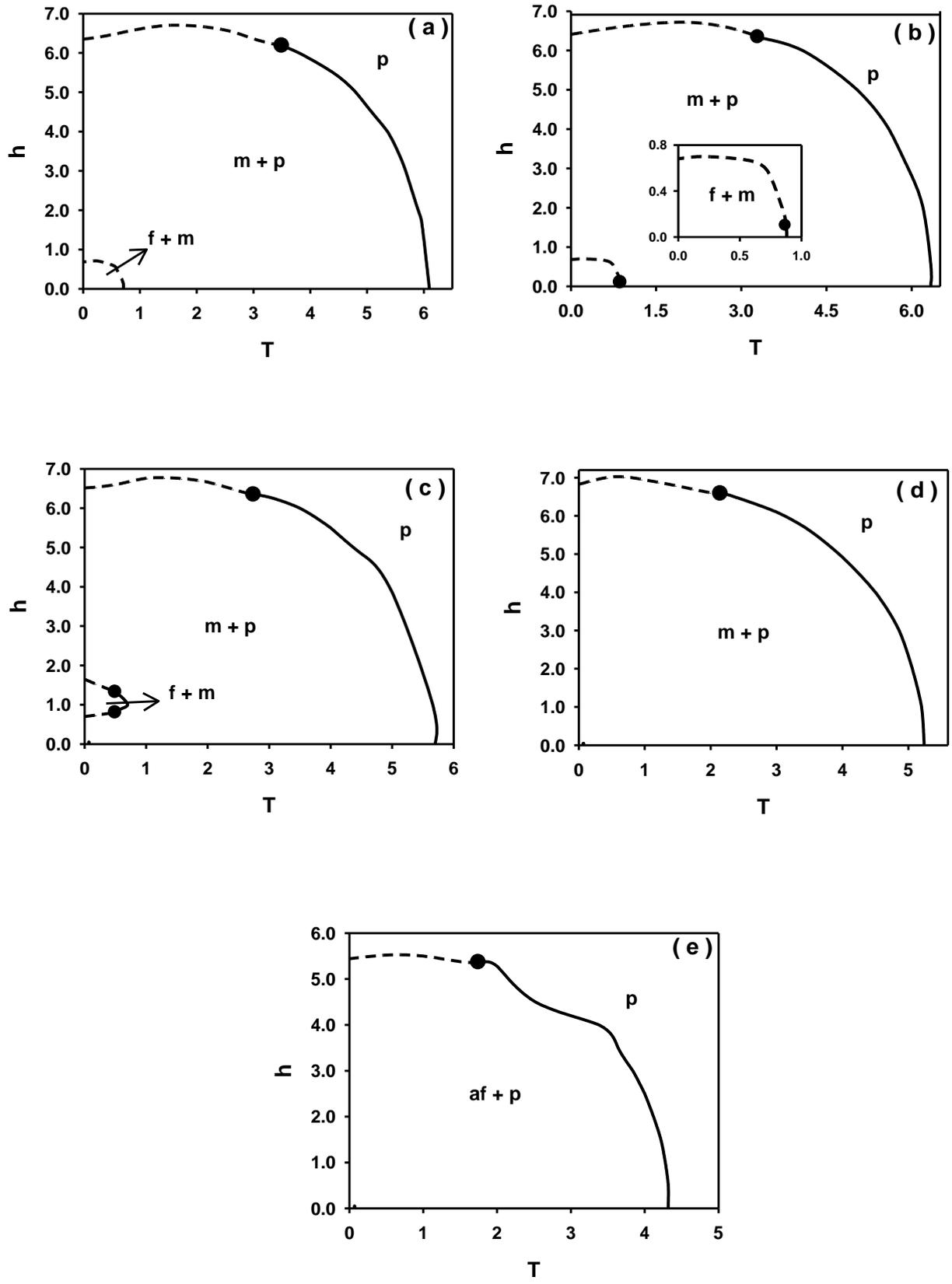

**Fig. 6**

| PHASES | Symbol | Configurations | Oscillation of Magnetizations ||||
|---|---|---|---|---|---|---|
|  |  |  | $m_1^A$ | $m_1^B$ | $m_2^A$ | $m_2^B$ |
| **Paramagnetic Phase** | p | 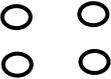 | 0 | 0 | 0 | 0 |
| **Ferromagnetic Phase** | f | 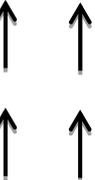 | 1 | 1 | 1 | 1 |
| **Antiferromagnetic Phase** | af | 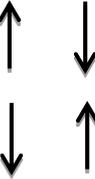 | 1 | -1 | -1 | 1 |
| **Surface Phase** | sf | 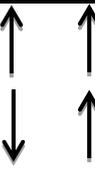 | 1 | 1 | -1 | 1 |
| **Compensated Phase** | c | 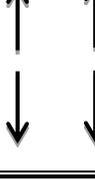 | 1 | 1 | -1 | -1 |
|  |  |  | -1 | -1 | 1 | 1 |
| **Mixed Phase** | m | 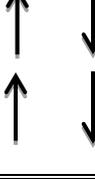 | 1 | -1 | 1 | -1 |
| **Non magnetic Phase** | nm | 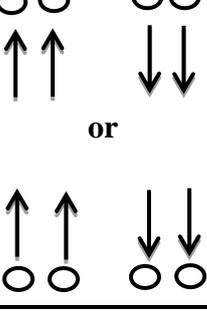 | 0 | 0 | 1 | 1 |
|  |  |  | 0 | 0 | -1 | -1 |
|  |  |  | 1 | 1 | 0 | 0 |
|  |  |  | -1 | -1 | 0 | 0 |

**Table 1**

ERCIYES UNIVERSITY
**DEPARTMENT OF PHYSICS**
**38039 KAYSERI-TURKEY**

PHONE: 90 (352) 207666 Ext: 33134
FAX: 90 (352) 4374933
E-Mail: mehmetertas@erciyes.edu.tr                November 23, 2014

Prof. Dr. Joel L. Lebowitz
Director, Center for Mathematical Sciences Research.
Rutgers, The State University
110 Frelinghuysen Road
Piscataway, NJ  08854-8019
Phone: 732-445-3117/3923
Fax: 732-445-4936
Email: lebowitz@math.rutgers.edu

Dear Prof. Dr. Lebowitz

I hereby respectfully submit a manuscript entitled "**The dynamic magnetic behaviors of the Blume-Capel Ising bilayer system**" for publication in **Journal of Statistical Physics**.

Looking forward to hearing from you soon.

Sincerely yours,

Associate Professor
Mehmet Ertaş

**Research Highlights**

- Dynamic behaviors in the spin-1 Blume-Capel Ising bilayer system is investigated.
- The dynamic phase transitions and dynamic phase diagrams are obtained.
- Dynamic phase diagrams are presented for FM/FM, AFM/FM and AFM/AFM interaction.
- Dynamic phase diagrams exhibit several ordered phases, coexistence phase regions as well as a re-entrant behavior.